\documentclass[journal,10pt,single]{IEEEtran}
\usepackage{amssymb}
\usepackage{amsfonts}
\usepackage{epstopdf}
\usepackage{graphicx}
\usepackage{stmaryrd}
\usepackage{amssymb,amsmath}
\usepackage{multirow,url}
\usepackage{color,cite}
\usepackage{hyperref}
\usepackage{comment}
\usepackage[OT1]{fontenc} 
\usepackage{subcaption}
\usepackage{diagbox} 

\ifodd 1
\newcommand{\com}[1]{\textbf{\color{red} (COMMENT: #1)}} 
\newcommand{\comg}[1]{\textbf{\color{green} (COMMENT: #1)}}
\newcommand{\response}[1]{\textbf{\color{magenta} (RESPONSE: #1)}} 
\else

\newcommand{\com}[1]{}
\newcommand{\comg}[1]{}
\newcommand{\response}[1]{}
\fi
\ifodd 0
\newcommand{\referred}[1]{\textcolor{red}{RefPaper: #1}} 
\else
\newcommand{\referred}[1]{}
\fi

\ifodd 0
\newcommand{\changeblue}[1]{\textcolor{blue}{Modified: #1}} 
\else
\newcommand{\changeblue}[1]{}
\fi

\begin{document}

\title{Information Update: TDMA or FDMA?}


\author{Haoyuan~Pan,~\IEEEmembership{Member,~IEEE,}，
       and~Soung~Chang~Liew,~\IEEEmembership{Fellow,~IEEE}
\thanks{H. Pan and S. C. Liew are with the Department of Information Engineering, The Chinese University of Hong Kong, Hong Kong.
Email: \{hypan, soung\}@ie.cuhk.edu.hk.
}
}

\maketitle 

\begin{abstract}
This paper studies information freshness in information update systems operated with TDMA and FDMA. Information freshness is characterized by a recently introduced metric, age of information (AoI), defined as the time elapsed since the generation of the last successfully received update. In an update system with multiple users sharing the same wireless channel to send updates to a common receiver, how to divide the channel among users affects information freshness. We investigate the AoI performances of two fundamental multiple access schemes, TDMA and FDMA. We first derive the time-averaged AoI by estimating the packet error rate of short update packets based on Gallager's random coding bound. For time-critical systems, we further define a new AoI metric, termed bounded AoI, which corresponds to an AoI threshold for the instantaneous AoI. Specifically, the instantaneous AoI is below the bounded AoI a large percentage of the time. We give a theoretical upper bound for bounded AoI. Our simulation results are consistent with our  theoretical analysis. Although TDMA outperforms FDMA in terms of average AoI, FDMA is more robust against varying channel conditions since it gives a more stable bounded AoI across different received powers. Overall, our findings give insight to the design of practical multiple access systems with AoI requirements.
\end{abstract}






\section{Introduction}

In recent years, Internet of Things (IoT) has enabled ubiquitous connection for a large number of devices with sensing, monitoring, and communication capabilities. The proliferation of IoT devices brings new applications demanding regular and frequent updates of certain information. For example, timely information updates are critical in real-time monitoring and control systems, such as air pollution monitoring in smart cities and motion control in industrial IoT \referred{IIoT,real_time_status,IoT_AoI_sample_update}\cite{IIoT,real_time_status,IoT_AoI_sample_update}.

Traditional communication networks focus on information rates and delays. With timely update requirements, the focus has been shifted towards the freshness of information\referred{AoI}\cite{AoI}. Information freshness is crucial to real-time systems since outdated information could potentially degrade the system performance. To measure information freshness, a metric termed age of information (AoI) was proposed\referred{AoI}\cite{AoI}. Specifically, AoI is age of the information last received by the receiver $-$ if at time $t$, the latest sample received by the receiver was a sample generated at the source at time $t - \tau $, then the instantaneous AoI at time $t$ is  $\tau $. Since AoI captures both generation time and delay, it is fundamentally different from traditional metrics such as packet loss rate and delay \referred{real_time_status,IoT_AoI_sample_update,AoI,AoIqueue,AoIscheduling1,AoIscheduling2}\cite{real_time_status,IoT_AoI_sample_update,AoI,AoIqueue,AoIscheduling1,AoIscheduling2}. 

In many wireless update systems, multiple users share the same wireless channel and send their latest samples to a common receiver. Multiple access schemes study how to divide the shared wireless channel among users \referred{tse2005fundamentals}\cite{tse2005fundamentals}. Although AoI has attracted considerable research interests, most prior work focused on upper layers, such as different systems modeled by various queueing models \referred{real_time_status,AoI,AoIqueue}\cite{real_time_status,AoI,AoIqueue}, and different AoI-aware packet scheduling policies \referred{AoIscheduling1,AoIscheduling2,IoT_AoI_sample_update}\cite{AoIscheduling1,AoIscheduling2,IoT_AoI_sample_update}. The impacts of different multiple access schemes on information freshness have not been well investigated, especially when packet error rates (PER) caused by unreliable channels at the PHY layer need to be taken into consideration. The main goal of this paper is to fill this gap. Specifically, we evaluate the AoI performances of two fundamental schemes, time-division multiple access (TDMA) and frequency-division multiple access (FDMA).

The major difference between TDMA and FDMA lies in their different ways of allocating resources. TDMA assigns orthogonal time slots to different users, while FDMA allocates nonoverlapping frequency bands to different users \referred{tse2005fundamentals}\cite{tse2005fundamentals}. Different time-frequency allocations affect AoI. A user in an FDMA system has longer transmission time due to the smaller allocated bandwidth, leading to possibly higher AoI because the information received was generated a longer time ago. TDMA, on the other hand, allows a user to send only in its own time slot, but it will take a longer time before the next update opportunity for the same user comes along. Adding to the above considerations are the different probabilities of decoding at the receiver for FDMA and TDMA. A quantitative study is needed to study the relative merits of the two systems when AoI is the performance metric. 

We consider two AoI metrics, average AoI and bounded AoI. Average AoI measures the time average of instantaneous AoI \referred{AoI,real_time_status}\cite{AoI,real_time_status}. Information update systems may require a low average AoI so that the received information over time is as fresh as possible. For many time-critical systems, however, it is not sufficient to have low average AoI. They require the instantaneous AoI to be upper bounded by a predefined threshold. More specifically, the percentage of time the instantaneous AoI is below the threshold should be larger than a target value $\gamma$. We define a new AoI metric, termed bounded AoI, to characterize the threshold given a target $\gamma$. For a given $\gamma$, lower bounded AoI means that the time-critical system can provide a higher level of information freshness.

In practice, packets containing update information from IoT are typically short (e.g., tens of bytes) \referred{IIoT}\cite{IIoT}. Shannon's capacity formula assumes infinite block length and is not adequate for characterization of the performance of short-packet systems. This paper uses Gallager's random coding bound (RCB) \referred{gallager}\cite{gallager} to estimate the packet error rate (PER) of short packets. With the help of RCB, we derive the theoretical average AoI of TDMA and FDMA systems. Furthermore, we give an upper bound of bounded AoI for each system. The upper bound analysis serves as a simple tool to compare the performance of TDMA and FDMA. 

We also validate the theoretical analysis by simulations. Our simulation results show when users have the same received power at the receiver, TDMA outperforms FDMA in terms of average AoI, but FDMA has a more stable bounded AoI than TDMA across different received powers. Since FDMA is more robust against varying channel conditions, our results also indicate that when users have different received powers, FDMA gives a more uniform bounded AoI performance across different users in multiuser time-critical information update systems. Overall, our findings provide insight into the design practical multiple access systems with AoI requirements.

\section{Backgrounds} \label{sec:backgrounds}
\subsection{Age of Information (AoI)} \label{sec:backgrounds1}
Throughout this paper, our performance metric is AoI. Suppose that there are totally $N$ users $\{1, 2, ... , n\}$. The $i$-th update packet from user $j$ is generated at time instant $t_i^j$ and is received by the receiver at time instant ${t_i^j}{'}$. For any given time instant $t$, the last received update from user $j$ is indexed by ${N_j}(t) = \max \{ i|{t_i^j}{'} \le t\}$. The timestamp of the most recently received update is ${U_j}(t) = {t_{{N_j}(t)}}$. Then, the instantaneous AoI of user $j$ is defined by
\begin{align}
{\Delta _j}(t) = t - {U_j}(t).
\end{align}

\noindent The smaller the ${\Delta _j}(t)$, the more updated the information from user $j$ is. We next present two AoI metrics, namely average AoI and bounded AoI.

\noindent\underline{\bf{Average AoI}}: Average AoI \referred{AoI}\cite{AoI} measures the time average of the instantaneous AoI ${\Delta _j}(t)$. The average AoI of user $j$ is 
\begin{align}
{\bar \Delta _j} = \mathop {\lim }\limits_{T \to \infty } \frac{1}{T}\int_0^T {{\Delta _j}(t)dt}. 
\end{align}

\noindent Considering the whole network, the average AoI among all the users is 
\begin{align}
\bar \Delta  = \frac{1}{N}\sum\limits_{j = 1}^N {{{\bar \Delta }_j}}.
\end{align}

\noindent\underline{\bf{Bounded AoI}}: For many time-critical systems, it is not sufficient to have low average AoI ${\bar \Delta _j}$; the instantaneous AoI ${\Delta _j}(t)$ needs to be upper bounded by a predefined threshold $\Delta _{THR}$. In practice, $\Delta _{THR}$ is determined by the system's timing requirement: a system may require that the percentage of time ${\Delta _j}(t)$ is below $\Delta _{THR}$ to be larger than or equal to a target value $\gamma$.

Specifically, given a target $\gamma$, we define bounded AoI $\Delta _{THR}$ as the smallest $\Delta _{THR}$ that satisfies 
\begin{align}
\Pr [{\Delta _j}(t) \le \Delta _{THR}] \ge \gamma ,\forall j \in \{ 1,2,...,n\},
\label{equ:bounded_aoi}
\end{align}

\noindent where $t$ is a random point in time. Given a target $\gamma$, we say that a system can provide a bounded AoI $\Delta _{THR}$ with confidence $\gamma$, if (\ref{equ:bounded_aoi}) is satisfied. In general, $\Delta _{THR}$ changes as $\gamma$ varies, i.e., $\Delta _{THR}$ is a function of $\gamma$. A lower $\Delta _{THR}$ means that the information update system can provide a higher level of information freshness.

We can use Chebyshev's inequality \referred{probability}\cite{probability} to estimate $\Delta _{THR}$. Specifically, let $\Delta _{THR[j]}$ denote the bounded AoI for user $j$. We compute the probability 
\begin{align}
\Pr [{\Delta _j}(t) &\le \Delta _{THR[j]}] = 1 - \Pr [{\Delta _j}(t) > \Delta _{THR[j]}] \notag\\ 
 &= {\rm{1}} - \Pr [{\Delta _j}(t) - {{\bar \Delta }_j}> \Delta _{THR[j]} - {{\bar \Delta }_j}] \notag\\
 &\ge 1 - \Pr [|{\Delta _j}(t) - {{\bar \Delta }_j}|> \Delta _{THR[j]} - {{\bar \Delta }_j}]  \label{equ:chebyshev2}\\
 & \ge 1 - \frac{{\sigma _j^2}}{{{{(\Delta _{THR[j]} - {{\bar \Delta }_j})}^2}}}
 \label{equ:chebyshev}
\end{align}

\noindent where (\ref{equ:chebyshev}) is obtained by Chebyshev's inequality from (\ref{equ:chebyshev2}). $\sigma _j^2$ is the variance of the instantaneous AoI over time
\begin{align}
\sigma _j^2 &= {\lim _{T \to \infty }}\frac{1}{T}\int_0^T {{{\left( {{\Delta _j}(t) - {{\bar \Delta }_j}} \right)}^2}} dt\notag\\ 
&= {\lim _{T \to \infty }}\frac{1}{T}\int_0^T {{\Delta _j}{{(t)}^2}} dt - {\left( {{{\bar \Delta }_j}} \right)^2}\notag\\ 
&= \overline {\Delta _j^2}  - {\left( {{{\bar \Delta }_j}} \right)^2}
\end{align}

\noindent where $\overline {\Delta _j^2}  = {\lim _{T \to \infty }}\frac{1}{T}\int_0^T {{\Delta _j}{{(t)}^2}} dt$. Given a fixed $\gamma$, we compute an upper bound $\widehat \Delta _{THR[j]}$ for the corresponding bounded AoI $\Delta _{THR[j]}$ for user $j$ by 
\begin{align}
\widehat \Delta _{THR[j]}= \sqrt {\frac{{\sigma _j^2}}{{1 - \gamma }}}  + {\bar \Delta _j} = \sqrt {\frac{{\overline {\Delta _j^2}  - {{\left( {{{\bar \Delta }_j}} \right)}^2}}}{{1 - \gamma }}}  + {\bar \Delta _j}.
\label{equ:bounded_aoi_upper_bounded}
\end{align}

\noindent Finally, since all users need to satisfy the time-critical requirement (\ref{equ:bounded_aoi}), we have an upper bound $\widehat \Delta _{THR}$ for $\Delta _{THR}$
\begin{align}
\widehat \Delta _{THR} = {\max _{j \in \{ 1,2,...,n\} }}\widehat \Delta _{THR[j]}.
\end{align}

In multiple access systems, the wireless channel is shared among multiple users. TDMA and FDMA are two common multiple access schemes on how to divide the channel among users. We analyze their AoI performances in Section \ref{sec:TDMA} and Section \ref{sec:FDMA}, respectively. Before the detailed analysis, we review Gallager's Random Coding Bound that will be used to estimate the packet error rate (PER). 

\subsection{Gallager's Random Coding Bound} \label{sec:backgrounds2}
In practical information update systems with AoI requirements, update packets are typically short. Information theory tells us that with finite block lengths, the PER cannot go to zero \referred{gallager}\cite{gallager}. In general, for a coded packet, the PER decreases as the coded block length increases. This paper uses the random coding bound (RCB) to estimate the PER. 

Gallager's RCB on the average block error probability $p$ of random $(L, K)$ codes has the form  $p \le {2^{ - L{E_G}(R)}}$ \referred{gallager}\cite{gallager}, where $K$ is the number of source bits of a packet; $L$ is the block length of coded bits; $R=K/L$ is the code rate. ${E_G}(R)$ is the random coding error exponent. Under perfect channel state information at the receiver, ${E_G}(R)$ is 
\begin{align}
{E_G}(R) = {\max _{0 \le \rho  \le 1}}[{E_0}(\rho ) - \rho R],
\label{equ:rcb}
\end{align}
\noindent where $\rho$ is the auxiliary variable over which optimization is performed to get the maximum value of the right-hand side of (\ref{equ:rcb}), and
\begin{align}
{E_0}(\rho ): =  - {\log _2}E\left[ {{{\left( {\frac{{E[{p_{Y|C}}{{(Y|C')}^{\frac{1}{{1 + \rho }}}}|Y]}}{{{p_{Y|C}}{{(Y|C)}^{\frac{1}{{1 + \rho }}}}}}} \right)}^\rho }} \right],
\end{align}
\noindent where $C$ and $C'$  are the two independent and uniformly distributed binary random variables, e.g., coded bits having values 0 or 1. $Y$ is the received signal in an AWGN channel. 

\section{AoI of TDMA Systems}\label{sec:TDMA}
This section analyzes the theoretical AoI performance of TDMA systems. Section \ref{sec:TDMA1}  first presents the TDMA system model. We consider symmetric channels in which all users have the same received power at the receiver. Section \ref{sec:TDMA2}  then derives the average AoI and bounded AoI for TDMA systems. 

\subsection{TDMA System Model}\label{sec:TDMA1}
\begin{figure}
\centering
\includegraphics[width=0.45\textwidth]{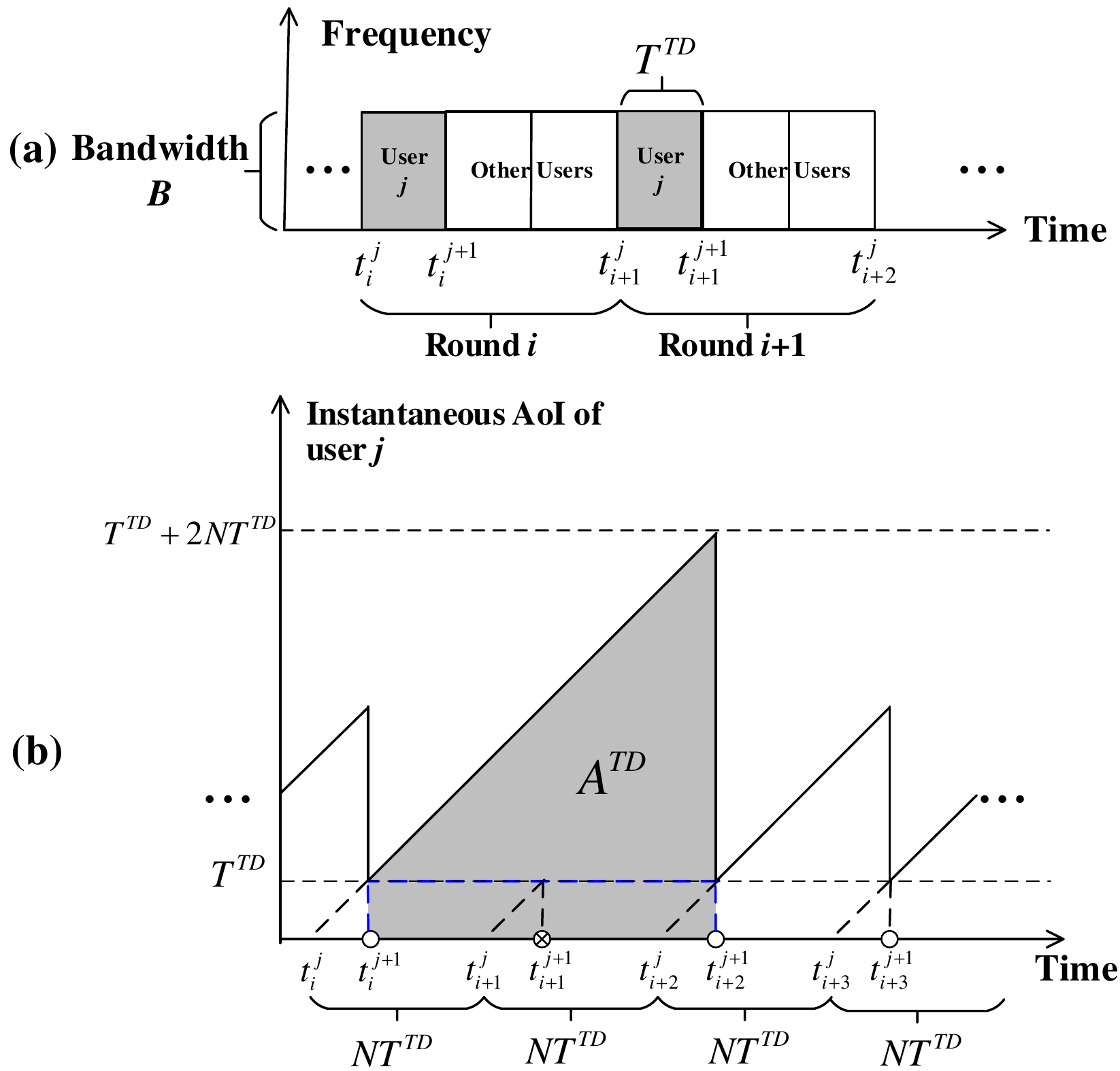}
\caption{(a) TDMA system model; (b) An example of instantaneous AoI for user $j$: three update packets are successfully decoded at $t_i^{j + 1}$, $t_{i + 2}^{j + 1}$, and $t_{i + 3}^{j + 1}$, while one packet fails to be decoded at $t_{i + 1}^{j + 1}$. }
\label{fig:tdma_model}
\end{figure}

In TDMA, time is divided into orthogonal time slots, which are then allocated to the different users. Suppose that different users send update packets to the receiver in a round-robin manner, as shown in Fig. \ref{fig:tdma_model}(a). Every user uses the whole bandwidth $B$ to transmit. We define a round as the total transmission time of all the $N$ users. Since with symmetric channels all users have the same received power $P$, we assume that each round consists of $N$ time slots and each user occupies one time slot with slot duration $T^{TD}$. The time duration of a round is $NT^{TD}$. 

In practical information update systems, physical observations are sensed and submitted at regular intervals. Suppose that in every round, each user generates a new update packet. Let interval $[{t_i},{t_{i + 1}}]$ denote the time period of round $i$, i.e., ${t_{i + 1}} = {t_i} + NT^{TD}$. In round $i$, user $j$ generates packet $C_i^j$ at $t_i^j$ and finishes transmission at $t_i^{j + 1} = t_i^j + T^{TD}$, where $t_i^j = {t_i} + (j - 1)T^{TD}$. In other words, each user sends packets at an interval $NT^{TD}$.

In round $i$, packet decoding for user $j$ occurs at $t_i^{j + 1}$. If $C_i^j$ is successfully decoded by the receiver, the instantaneous AoI ${\Delta _j}(t)$ is reset to $T^{TD}$. Fig. \ref{fig:tdma_model}(b) shows an example of ${\Delta _j}(t)$ in an $N$-user TDMA system, where ${\Delta _j}(t)$  drops to $T^{TD}$ at $t_i^{j + 1}$, $t_{i + 2}^{j + 1}$, and $t_{i + 3}^{j + 1}$ because of successful packet decodings. One packet fails to be decoded at $t_{i + 1}^{j + 1}$, so ${\Delta _j}(t)$ continues to increase linearly.

\subsection{Average AoI and Bounded AoI of TDMA Systems}\label{sec:TDMA2}
We now derive the average AoI $\bar \Delta^{TD}$ and estimate the bounded AoI $\widehat \Delta _{THR}^{TD}$ for an $N$-user TDMA system. 

\noindent\underline{\bf{Average AoI}}: Suppose that user $j$'s packet $C_i^j$ is successfully decoded with probability $1 - {p^{TD}}$, where ${p^{TD}} \in [0,1]$ denotes the PER for TDMA. After one successful update, denote by $X$ the number of packets transmitted until the next successful update. Then $X$ is a geometric random variable with a probability mass function
\begin{align}
{P_X}(X = x) = {({p^{TD}})^{x - 1}}(1 - {p^{TD}}),x = 1,2,...
\label{equ:X_distribution}
\end{align}

\noindent i.e., $x-1$ failed decodings followed by one successful decoding.\footnote{Note that $E[X] = \frac{1}{{1 - {p^{TD}}}}$, $E[{X^2}] = \frac{{1 + {p^{TD}}}}{{{{(1 - {p^{TD}})}^2}}}$, and $E[{X^3}] = 1 + \frac{{7{p^{TD}}}}{{1 - {p^{TD}}}} - \frac{{12{{({p^{TD}})}^2}}}{{{{(1 - {p^{TD}})}^2}}} + \frac{{6{{({p^{TD}})}^3}}}{{{{(1 - {p^{TD}})}^3}}}$.}

To compute the average AoI $\bar \Delta _j^{TD}$ of user $j$, let us consider the area ${A^{TD}}$ (see Fig. \ref{fig:tdma_model}(b)) between two consecutive successful updates. Suppose that the last successful update occurs at $t_i^{j + 1}$ and the next successful update occurs at $t_{i + X}^{j + 1}$. The time elapsed is $t_{i + X}^{j + 1} - t_i^{j + 1} = XNT_{}^{TD}$, i.e., the time duration of $X$ rounds.   ${A^{TD}}$ is computed by
\begin{align}
{A^{TD}} &= \int_{t_i^{j + 1} }^{t_{i + X}^{j + 1}} {{\Delta _j}(t)} dt = XN{T^{TD}} \cdot {T^{TD}} + \frac{(XN{T^{TD}})^2}{2}.
\end{align}


\noindent The time slot duration $T^{TD}$ is related to the block length ${L^{TD}}$ and the system bandwidth  $B$, i.e., $T^{TD}$ = $\frac{{{L^{TD}}}}{B}$. Thus, $\bar \Delta _j^{TD}$ can be computed by (\ref{equ:average_aoi_td}). It follows immediately that for symmetric channels, the average AoI of the network is $\bar \Delta ^{TD}$ = $\frac{1}{N}\sum\limits_{j = 1}^N {\bar \Delta _j^{TD}}$  = $\bar \Delta _j^{TD}$. We will drop the index $j$ since individual users and the network have the same average AoI (as well as bounded AoI) under symmetric channels.
\newcounter{mytempeqncnt}
\begin{figure*}[!t]
\setcounter{mytempeqncnt}{\value{equation}}
\footnotesize
\begin{align}
\bar \Delta _j^{TD} = \mathop {\lim }\limits_{W \to \infty } \frac{{\sum\limits_{w = 1}^W {A_w^{TD}} }}{{\sum\limits_{w = 1}^W {{X_w}N{T^{TD}}} }} = \mathop {\lim }\limits_{W \to \infty } \frac{{\sum\limits_{w = 1}^W {{X_w}N{T^{TD}} \cdot {T^{TD}} + \frac{{{{({X_w}N{T^{TD}})}^2}}}{2}} }}{{\sum\limits_{w = 1}^W {{X_w}N{T^{TD}}} }} = {T^{TD}} + \frac{{N{T^{TD}}E[{X^2}]}}{{2E[X]}}
= \left( {1 + \frac{N}{{1 - {p^{TD}}}} - \frac{N}{2}} \right)\frac{{{L^{TD}}}}{B}
\label{equ:average_aoi_td}
\end{align}

\begin{align}
{\overline {\Delta _j^2} ^{TD}} &= {\lim _{T \to \infty }}\frac{1}{T}\int_0^T {{\Delta _j}{{(t)}^2}} dt = \mathop {\lim }\limits_{W \to \infty } \frac{{\sum\limits_{w = 1}^W {\int_{t_i^{j + 1} }^{t_{i + X}^{j + 1} } {{\Delta _j}{{(t)}^2}} dt} }}{{\sum\limits_{w = 1}^W {{X_w}N{T^{TD}}} }}
 = \mathop {\lim }\limits_{W \to \infty } \frac{{\sum\limits_{w = 1}^W {\frac{{{{({X_w})}^3}{N^3}{{({T^{TD}})}^3}}}{3} + {{({X_w})}^2}{N^2}{{({T^{TD}})}^3} + {X_w}N{{({T^{TD}})}^3}} }}{{\sum\limits_{w = 1}^W {{X_w}N{T^{TD}}} }} \notag \\
&= \left( {\frac{{{N^2}E[{X^3}]}}{{3E[X]}} + \frac{{NE[{X^2}]}}{{E[X]}} + 1} \right) \cdot {\left( {\frac{{{L^{TD}}}}{B}} \right)^2}
\label{equ:second_moment_td}
\end{align}
\hrulefill
\end{figure*}

\noindent\underline{\bf{Bounded AoI}}: We first estimate the bounded AoI $\Delta _{THR[j]}^{TD}$ for user $j$. ${\overline {\Delta _j^2} ^{TD}}$ is computed by (\ref{equ:second_moment_td}). Substituting  (\ref{equ:average_aoi_td}) and (\ref{equ:second_moment_td}) into (\ref{equ:bounded_aoi_upper_bounded}), we obtain an upper bound $\widehat \Delta _{THR[j]}^{TD}$ of $\Delta _{THR[j]}^{TD}$. With symmetric channels, $\widehat \Delta _{THR}^{TD}$ = $\widehat \Delta _{THR[j]}^{TD}$. We will present the simulation results of  $\bar \Delta^{TD}$ and $\widehat \Delta _{THR}^{TD}$ in Section \ref{sec:simulation}.

\begin{figure}
\centering
\includegraphics[width=0.45\textwidth]{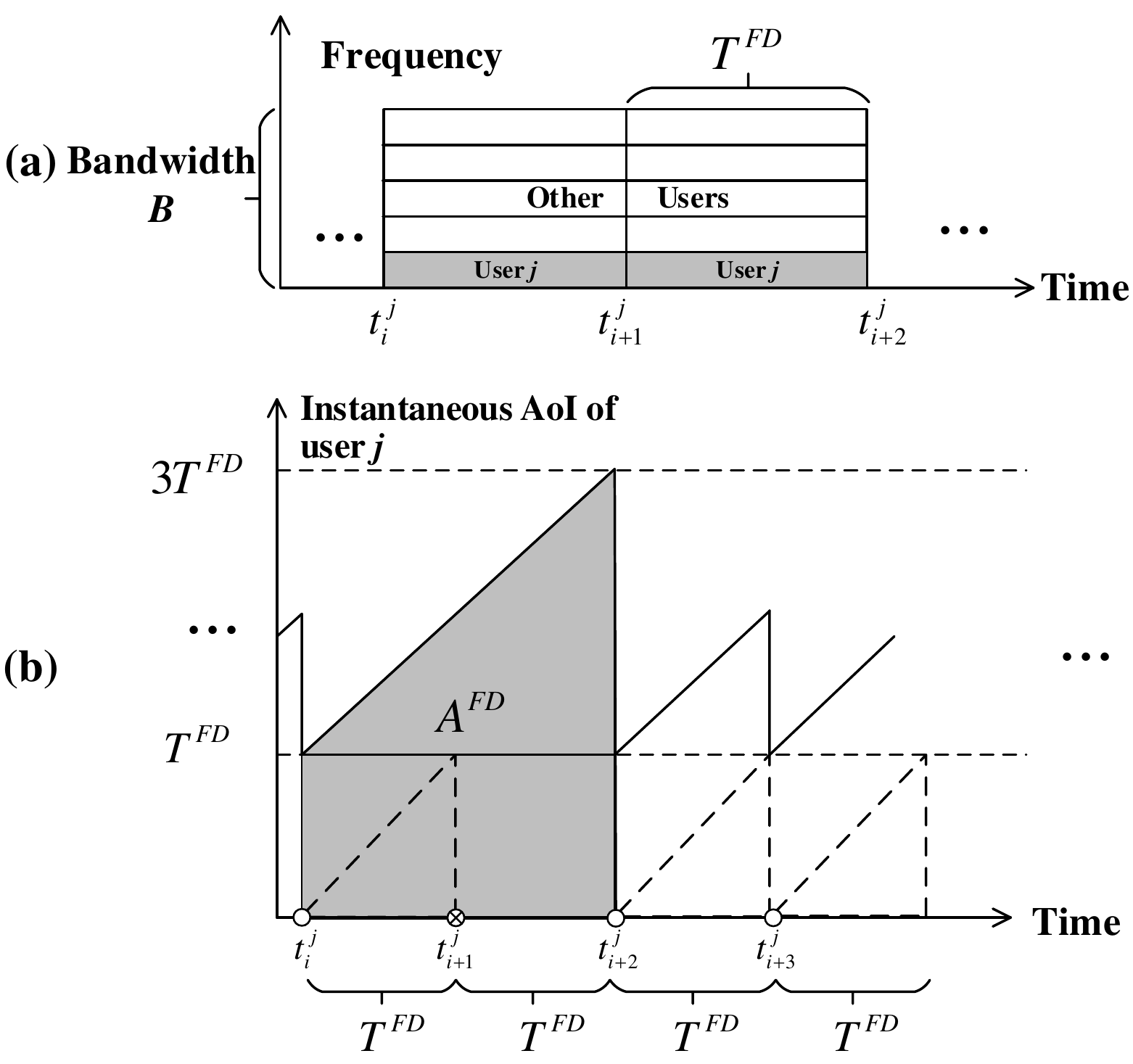}
\caption{(a) FDMA system model; (b) An example of instantaneous AoI for user $j$:three update packets are successfully decoded at $t_i^j$, $t_{i + 2}^j$, and $t_{i + 3}^j$, while one packet fails to be decoded at $t_{i + 1}^j$. }
\label{fig:fdma_model}
\end{figure}

\section{AoI of FDMA Systems}\label{sec:FDMA}
Let us now focus on FDMA. Section \ref{sec:FDMA1} presents the system model and Section \ref{sec:FDMA2} derives the average AoI and bounded AoI. We consider symmetric channels in which all users have the same received power at the receiver and are allocated equal bandwidths. 

\subsection{FDMA System Model}\label{sec:FDMA1}
In FDMA, the bandwidth is divided into nonoverlapping channels and each user is then assigned a different channel for multiple access. In other words, all the users send at the same time but in different frequency bands, as shown in Fig. \ref{fig:fdma_model}(a). 

By symmetric channels, we assume that all users have the same received power $P$ and are allocated the same bandwidth $\frac{B}{N}$. Let $T_{}^{FD}$ denote the packet duration. Since FDMA users are not affected by each other, a user can continuously send packets to the receiver at an interval $T^{FD}$. As shown in Fig. \ref{fig:fdma_model}(a), user $j$ generates packet $C_i^j$ at $t_i^j$  and finishes transmission at $t_{i + 1}^j = t_i^j + {T^{FD}}$. The packet decoding for $C_i^j$ occurs at $t_{i + 1}^j$. Fig. \ref{fig:fdma_model}(b) shows an example of the instantaneous AoI ${\Delta _j}(t)$ of user $j$ in an $N$-user FDMA system: for successful decodings, ${\Delta _j}(t)$ is reset to $T^{FD}$ at  $t_i^j$, $t_{i + 2}^j$, and $t_{i + 3}^j$; one packet fails to be decoded at $t_{i + 1}^j$. 

\subsection{Average AoI and Bounded AoI of FDMA Systems}\label{sec:FDMA2}

We now derive the average AoI $\bar \Delta^{FD}$  and estimate the bounded AoI $\widehat \Delta _{THR}^{FD}$ for an $N$-user FDMA system. 

\noindent\underline{\bf{Average AoI}}: Suppose that packet $C_i^j$ is successfully decoded with probability $1 - {p^{FD}}$, where ${p^{FD}} \in [0,1]$ denotes the PER for FDMA. After one successful update, the number of packets transmitted until the next successful update $X$ has the same probability mass function as (\ref{equ:X_distribution}), except that ${p^{TD}}$ is replaced by ${p^{FD}}$.

We follow the same method as in TDMA to compute the average AoI $\bar \Delta _j^{FD}$ for user $j$. Specifically, we compute the area ${A^{FD}}$ (see Fig. \ref{fig:fdma_model}(b)) between two consecutive successful updates at $t_i^j$ and $t_{i + X}^j$, and the time elapsed is $t_{i + X}^j - t_i^j$ = $X{T^{FD}}$. Let ${L^{FD}}$ denote the block length. Then ${T^{FD}}$ is $\frac{{N{L^{FD}}}}{B}$ due to equal bandwidth allocation. With symmetric channels, $\bar \Delta^{FD}$ is the same as $\bar \Delta _j^{FD}$ (\ref{equ:average_aoi_fd}). 


\noindent\underline{\bf{Bounded AoI}}: For FDMA, to estimate the bounded AoI $\Delta _{THR[j]}^{FD}$, ${\overline {\Delta _j^2} ^{FD}}$ is computed by (\label{equ:second_moment_fd}). Substituting (\ref{equ:average_aoi_fd}) and (\ref{equ:second_moment_fd}) into (\ref{equ:bounded_aoi_upper_bounded}), we obtain an upper bound $\widehat \Delta _{THR[j]}^{FD}$ of $\Delta _{THR[j]}^{FD}$, and $\widehat \Delta _{THR}^{FD}$ = $\widehat \Delta _{THR[j]}^{FD}$. The next section compares the AoI performances between TDMA and FDMA systems.
%


\begin{figure*}[!t]
\setcounter{mytempeqncnt}{\value{equation}}
\footnotesize
\begin{align}
\bar \Delta ^{FD} = \bar \Delta _j^{FD} = \mathop {\lim }\limits_{W \to \infty } \frac{{\sum\limits_{w = 1}^W {A_w^{FD}} }}{{\sum\limits_{w = 1}^W {{X_w}{T^{FD}}} }} = \mathop {\lim }\limits_{W \to \infty } \frac{{\sum\limits_{w = 1}^W {{X_w}{T^{FD}} \cdot {T^{FD}} + \frac{{{{({X_w}{T^{FD}})}^2}}}{2}} }}{{\sum\limits_{w = 1}^W {{X_w}{T^{FD}}} }} = {T^{FD}} + \frac{{{T^{FD}}E[{X^2}]}}{{2E[X]}}
= \left( {\frac{1}{2} + \frac{1}{{1 - {p^{FD}}}}} \right)\frac{{N{L^{FD}}}}{B}
\label{equ:average_aoi_fd}
\end{align}
\begin{align}
{\overline {\Delta _j^2} ^{FD}} &= {\lim _{T \to \infty }}\frac{1}{T}\int_0^T {{\Delta _j}{{(t)}^2}} dt = \mathop {\lim }\limits_{W \to \infty } \frac{{\sum\limits_{w = 1}^W {\int_{t_i^j }^{t_{i + X}^j} {{\Delta _j}{{(t)}^2}} dt} }}{{\sum\limits_{w = 1}^W {{X_w}{T^{FD}}} }}
= \mathop {\lim }\limits_{W \to \infty } \frac{{\sum\limits_{w = 1}^W {\frac{{{{({X_w})}^3}{{({T^{FD}})}^3}}}{3} + {{({X_w})}^2}{{({T^{FD}})}^3} + {X_w}{{({T^{FD}})}^3}} }}{{\sum\limits_{w = 1}^W {{X_w}{T^{FD}}} }} \notag \\
&= \left( {\frac{{E[{X^3}]}}{{3E[X]}} + \frac{{E[{X^2}]}}{{E[X]}} + 1} \right) \cdot {\left( {\frac{{N{L^{FD}}}}{B}} \right)^2}
\label{equ:second_moment_fd}
\end{align}
\hrulefill
\end{figure*}

\section{AoI Performance Comparison of TDMA and FDMA Systems}\label{sec:simulation}
We now evaluate the AoI performances of TDMA and FDMA systems. We verify the theoretical analysis in Sections \ref{sec:TDMA} and \ref{sec:FDMA} via simulations. In our simulations, we use an LDPC code in which a fixed number of four bits in a source packet are randomly selected and added to generate a coded bit (i.e., a fixed degree 4). The number of source bits per packet is $K=100$ and the block length of an LDPC-coded packet varies from 100 to 400 bits. BP decoding is used to decode the LDPC code \referred{tse2005fundamentals}\cite{tse2005fundamentals}.

\begin{figure}
\centering
\includegraphics[width=0.5\textwidth]{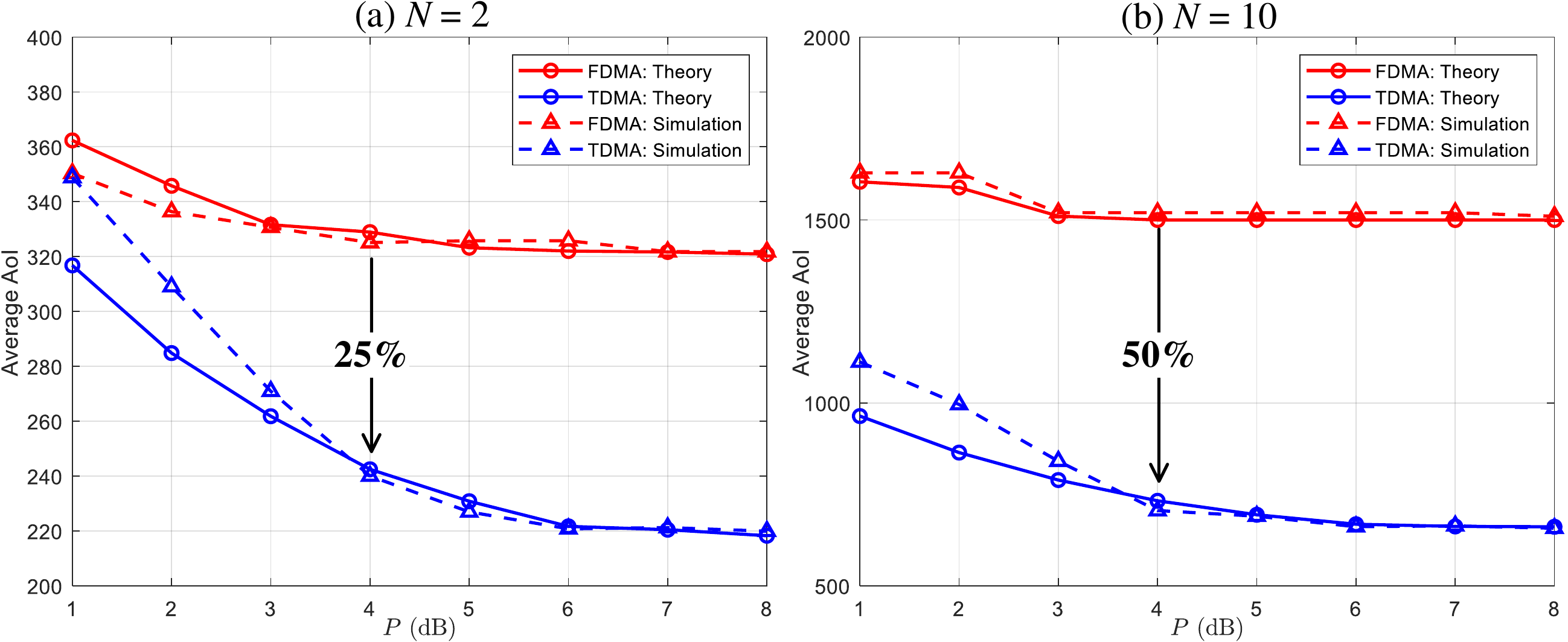}
\caption{Average AoI of TDMA and FDMA with (a) two and (b) ten users. A source packet has 100 bits. For each $P$, the average AoI is minimized by optimizing the block lengths from 100 to 400.}
\label{fig:average_aoi}
\end{figure}

\noindent\underline{\bf{Average AoI}}:  Fig. \ref{fig:average_aoi} plots the average AoI $\bar \Delta^{TD}$ and $\bar \Delta^{FD}$ versus the received power $P$, when the number of users $N$ is (a) 2 and (b) 10. We normalize both the bandwidth and the noise power to 1 so that $P$ has a unit of dB. For each $P$, we optimize the block lengths ${L^{TD}}$ and ${L^{FD}}$ to minimize $\bar \Delta^{TD}$ and $\bar \Delta^{FD}$, respectively. We can see from Fig. \ref{fig:average_aoi} that the simulation results are consistent with the theoretical analysis. In particular, TDMA has a significantly lower average AoI than that of FDMA. For example, TDMA reduces the average AoI by 50\% compared with FDMA, when $N=10$ and $P=4dB$. 

To explain the performance gain of TDMA over FDMA, let us assume that $P$ is high enough such that both PERs ${p^{TD}}$ and ${p^{FD}}$ approach zero. We have $\bar \Delta^{TD} = \left( {1 + \frac{N}{2}} \right)\frac{{{L^{TD}}}}{B}$ and $\bar \Delta^{FD} = \left( {\frac{{3N}}{2}} \right)\frac{{{L^{FD}}}}{B}$. It is easy to see that $\bar \Delta^{TD}<\bar \Delta^{FD}$ when $N>1$ and ${L^{TD}}$ = ${L^{FD}}$. This is because when $N$ is large, FDMA has a much longer packet duration ${T^{FD}}{\rm{ = }}\frac{{N{L^{FD}}}}{B}$. Although an FDMA user can continuously send packets to the receiver, the instantaneous AoI can only drop to ${T^{FD}}$ for a successful update. By contrast, TDMA drops the instantaneous AoI to ${T^{TD}}{\rm{ = }}\frac{{{L^{TD}}}}{B}$, thus having lower average AoI.

\begin{figure}
\centering
\includegraphics[width=0.5\textwidth]{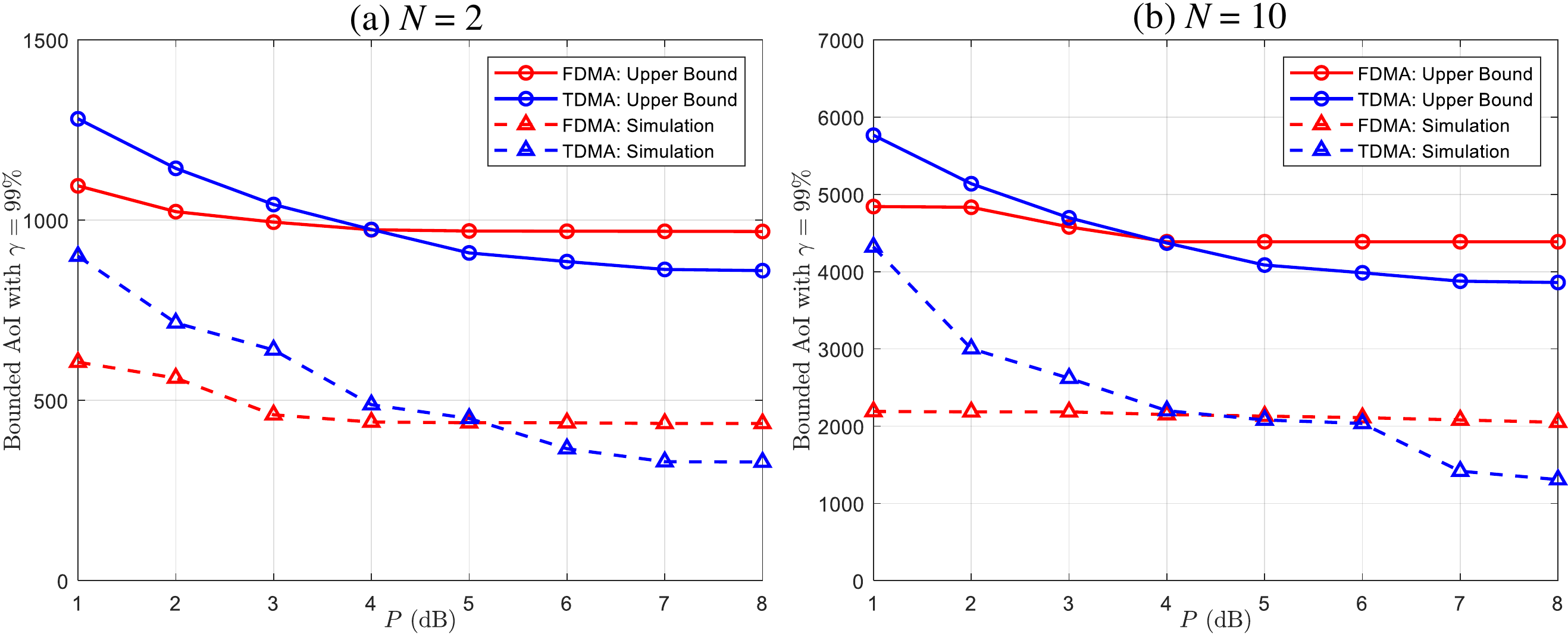}
\caption{Bounded AoI of TDMA and FDMA with (a) two and (b) ten users. A source packet has 100 bits. For each $P$, the bounded AoI is minimized by optimizing the block lengths from 100 to 400.}
\label{fig:bounded_aoi}
\end{figure}

\noindent\underline{\bf{Bounded AoI}}: We next fix $\gamma$=99\% to compare the bounded AoI when $N$ is (a) 2 and (b) 10. Fig. \ref{fig:bounded_aoi} plots the simulated bounded AoI $\Delta _{THR}^{TD}$ and $\Delta _{THR}^{FD}$, as well as the theoretical upper bounds $\widehat \Delta _{THR}^{TD}$ and $\widehat \Delta _{THR}^{FD}$  derived in Sections \ref{sec:TDMA} and \ref{sec:FDMA}. We see from Fig. \ref{fig:bounded_aoi} that there is a gap between $\Delta _{THR}^{TD}(\Delta _{THR}^{FD})$ and $\widehat \Delta _{THR}^{TD}(\widehat \Delta _{THR}^{FD})$, which we believe is due to the non-tightness of Chebyshev's inequality \referred{probability}\cite{probability}. Despite the gaps, both simulations and theoretical analysis indicate that FDMA has a lower bounded AoI when the received power $P$ is small, and TDMA gradually outperforms FDMA as $P$ increases. We explain this result as follows 

When $P$ is high, $\Delta _{THR}^{TD}$ is lower than $\Delta _{THR}^{FD}$. Assuming $P$ is high enough such that both ${p^{TD}}$ and ${p^{FD}}$ approach zero, it is easy to verify that TDMA and FDMA have the same variance $\frac{{{N^2}{L^2}}}{{12{B^2}}}$ if  ${L^{TD}} = {L^{FD}} = L$. As a result, the bounded AoI depends more on the average AoI when $P$ is high (see (\ref{equ:bounded_aoi})). That is, $\Delta _{THR}^{TD} < \Delta _{THR}^{FD}$  since $\bar \Delta ^{TD} < \bar \Delta^{FD}$ (see Fig. \ref{fig:average_aoi}). 

When $P$ is low, $\Delta _{THR}^{FD}$ is lower than $\Delta _{THR}^{TD}$. This is because TDMA has a larger variance of the instantaneous AoI. Specifically, for the same $P$, TDMA has a lower effective SNR than that of FDMA (i.e., $\frac{P}{B}$ vs. $\frac{{NP}}{B}$) and has a higher PER ${p^{TD}}$ than ${p^{FD}}$. Therefore, TDMA has a large variance of $X$, where $X$ is the number of packets transmitted between two successful updates. This means that with TDMA, the instantaneous AoI fluctuates more significantly, e.g., it drops to ${T^{TD}}$ after an update and is increased by $XN{T^{TD}}$  until the next update. By contrast, thanks to a smaller variance of $X$, FDMA has lower bounded AoI when $P$ is small. 

In summary, we observe from Fig. \ref{fig:bounded_aoi} that $\Delta _{THR}^{FD}$ is more stable than $\Delta _{THR}^{TD}$ across different received powers, indicating that FDMA is more robust than TDMA against varying channel conditions. This is important to a multiuser time-critical information update system consisting of a mix of high-SNR users and low-SNR users at the same time. Instead of frequently optimizing the time slot durations for different users according to their time-varying received SNRs in TDMA, FDMA can simply allocate equal bandwidth to different users. This greatly simplifies the system design but provides a more uniform bounded AoI performance across different users. 

\section{Conclusions}\label{sec:Conclusions}

We have compared the AoI performances between two basic multiple access schemes, TDMA and FDMA. Using Gallager's random coding bound to estimate the packet error rate of short packets, we derive the theoretical time-averaged AoI for TDMA and FDMA. Furthermore, we define a new AoI metric, bounded AoI, for time-critical systems. We give an upper bound of bounded AoI. 

Our simulation results show that, when all users have the same received power, TDMA has a lower average AoI than FDMA does. However, FDMA has a more stable bounded AoI across different received powers, indicating that FDMA is more robust than TDMA against varying channel conditions. This also means that when users have different received powers, FDMA can give a more uniform bounded AoI performance across different users. 

\bibliographystyle{IEEEtran}
\bibliography{aoi_tdma_fdma}

\end{document}